\documentclass[english,final,authoryear,5p,times,twocolumn]{elsarticle}

\usepackage{natbib}
\usepackage[utf8]{inputenc}
\usepackage{times}
\usepackage{gensymb}
\usepackage{color}

\begin{document}

\begin{frontmatter}

\title{Spectral actinometry at SMEAR-Estonia.}

\author[TO]{Andres Kuusk\corref{cor1}}
\author[TO]{Joel Kuusk}

\address[TO]{Tartu Observatory, University of Tartu, 61602 T\~{o}ravere, 
Estonia}
\cortext[cor1]{Corresponding author.
Tel.: +372 737 4528; fax: +372 737 4555
\\ {\it\hspace*{1.5em} E-mail address:} Andres.Kuusk@to.ee (A. Kuusk).}

\begin{abstract}

Systematic spectral measurements of downwelling solar radiation, both of
global and diffuse, have been collected in summertime during 8 years in
the hemi-boreal zone in south east Estonia near the SMEAR-Estonia
research station. The measurements provided information about the
variation of spectral and total, global and diffuse irradiance in the
wavelength range from 300 to 2160~nm with spectral resolution of 3~nm in
UV to 16~nm in SWIR spectral regions. Unique data have been collected
and quantitative description of the variability of the measured spectra
is provided. For the description of the synoptic situation during
measurements, instead of cloud cover, the ratio of measured to possible
total irradiance in the spectral range of 320--1800~nm is used. This
ratio could be used as the primary meteorological parameter instead of
cloud cover which is difficult to measure instrumentally.

\end{abstract}


\end{frontmatter}


\section{Introduction}
\label{intro}

Measurements of global and diffuse downwelling radiation are carried out
in Estonia at Tartu-Tõravere actinometric station of Estonian Weather
Service since 1955 \citep{Russak2003}. Both, global and diffuse
irradiance are measured with pyranometers which are spectrally
integrating sensors. While in the studies of energy fluxes in the
atmosphere and at Earth's surface global fluxes are needed, in
ecological studies of the vegetation-atmosphere interaction the energy
distribution over the radiation spectrum is important. \citet{John2013}
studied the allometry of cells and tissues within leaves, and found that
future work is needed to consider the possible influence of the
environment. Many trait allometries shift substantially due to
plasticity across different growing conditions, i.e., different supplies
of light, nutrients, and/or water. The study by \citet{Solomakhin2010}
showed how the changes in the spectrum of incident light due to colored
hailnets affect leaf anatomy, vegetative and reproductive growth as well
as fruit coloration in apple. \citet{Ji2020} found that solar radiation
components photosynthetically active radiation and \mbox{ultraviolet‐B}
have different associations with leaf nitrogen and phosphorus content.
The direct associations, when solar radiation is indicated by spectral
components, are greater than the indirect associations. So when
predicting the effects of global dimming on ecosystem nutrient fluxes,
the roles of direct, diffuse, and spectral components of solar
irradiance must be distinguished. The study by \citet{Moon2020}
concludes that accurate and spectrally resolved canopy radiative
transfer models are critically necessary to realistically determine
chemical reactions and gas concentrations within plant canopies and in
the immediately overlying atmospheric boundary layer.

The spectral composition of extraterrestrial solar radiation is
monitored on-board of satellites \citep{Harder2005,Harder2010,Kopp2014}.
For designing photoelectric solar power equipment some episodic
measurements of irradiance spectra at Earth's surface have been carried
out \citep{Norton2015}, and model simulations with atmospheric radiative
transfer models for standard situations are provided
\citep{Solar1.5,Gueymard2004}.  Eddy covariance (EC) sites for measuring
fluxes of trace gases usually include optical sensors measuring
downwelling, and sometimes also upwelling shortwave radiation, however,
a review by \citet{Balzarolo2011} revealed that only 5 out of 40
European EC sites involved in the study had hyperspectral radiometers
while the majority used multispectral or broadband sensors.  However, as
\citet{Ruizarias2018} state, irradiance data of better accuracy and with
better spatio-temporal resolution are needed both for the validation of
clear-sky solar radiation models and for solar energy industry.

In 2013--2015 a Station for Measuring Ecosystem-Atmosphere Relations,
SMEAR-Estonia was established in south east Estonia at the Järvselja
Experimental Forestry station \citep{SMEARE2015}. The station completes
the network of Finnish SMEAR-stations \citep{SMEARE2015}.  The
coordinates of the SMEAR-Estonia are (58\degree\ 16'~6"~N, 27\degree\
18'~16"~E).  For providing the research program of the SMEAR station
with radiation data a special spectrometer SkySpec was designed.
Measurements of global and diffuse solar irradiance with SkySpec started
at SMEAR-Estonia in August 2013. Measurements are carried out during
vegetation period.

\section{SkySpec spectrometer}
\label{spectrometer}

SkySpec spectrometer is a purpose-built instrument for continuous
measurement of global and diffuse downwelling spectral irradiance in the
shortwave spectral domain covering with three spectrometer modules the
wavelength range of 295--2205~nm.  Spectral resolution is 3~nm in
ultraviolet (UV), 10~nm in visible and near infrared (VNIR), and 16~nm
in shortwave infrared (SWIR) spectral region. The spectral range is
covered by 541 spectral bands. The standard scenario used for field
measurements includes the following stages: dark signal (1~minute),
global irradiance \mbox{(5~minutes)}, diffuse irradiance (1~minute),
global irradiance \mbox{(5~minutes)}. This scenario is repeated in
continuous loop when the solar zenith angle SZA $< 90\degree$. For
90$\degree <$ SZA $< 95\degree$ diffuse measurement stage is omitted. No
measurements are done if SZA $> 95\degree$. If irradiance levels are
very low and integration time of any of the spectrometer modules exceeds
the duration of current measurement stage, the stage is extended and at
least one measurement is acquired with all three spectrometer modules.
Integration time is automatically adjusted independently for each
spectrometer module so that the maximal value of the recorded raw signal
is between 60\% and 90\% of the full scale value.  Maximal usable
integration times are limited by dark signal levels and are set to
200~s, 250~s, and 1.5~s for the UV, VNIR, and SWIR modules of the
spectrometer, respectively. Dark measurements are made by covering the
fore-optics with a mechanical shutter. A shadow disk is used for
blocking approximately 6$\degree$\ in the direction of the Sun during
the diffuse measurements. An azimuth-elevation tracking system is used
for moving the shadow disk with complete fore-optics tracking the Sun in
azimuthal direction.  The detailed description of the spectrometer,
calibration and metrological processing of data is provided by
\citet{Kuusk2018a}. The spectrometer was radiometrically calibrated once
a year in Tartu Observatory.

\begin{figure}[t!]
\centering
\includegraphics[width=.9\columnwidth,
natwidth=2103,natheight=1464]{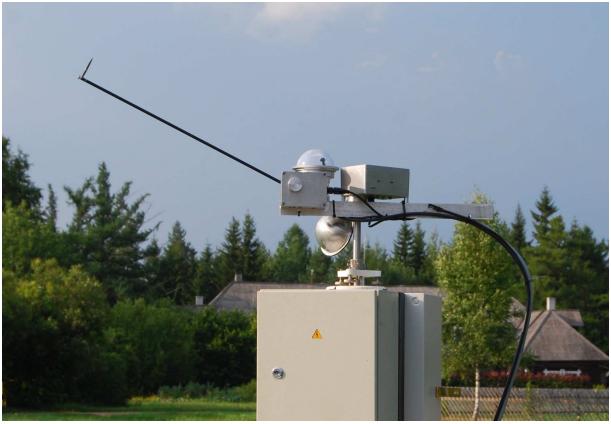}
\caption{SkySpec at Järvselja, Estonia measuring diffuse sky irradiance.}
\label{skyspec_a}
\end{figure}

\section{Data}
\label{data}

\begin{figure}[!ht]
\centering
\resizebox*{0.6\columnwidth}{!}{\includegraphics{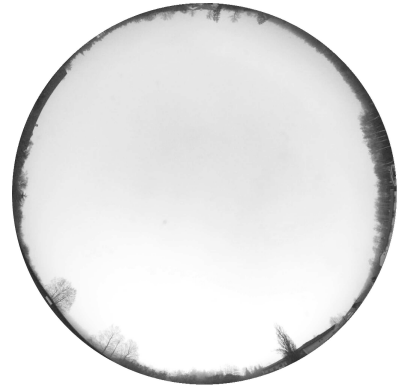}}
\caption{The cover of horizon at the spectrometer in 2013--2016.}
\label{horisont}
\end{figure}

The summary of SkySpec measurements at Järvselja is in Table
\ref{dates}. From 2013 to 2016 the spectrometer was installed on an open
place on ground at the distance of 1~km from the SMEAR-Estonia. The
cover of horizon at the spectrometer is shown in Fig. \ref{horisont}.
Since 2017 the SkySpec is installed near the SMEAR-Estonia atop a 30-m
tower at the distance of 45~m from the SMEAR tower. Thus the horizon is
open in all directions except the 130~m tower covers low sun for a while
in spring and autumn.

\begin{table}[ht!]
\centering
\caption{Summary of SkySpec measurements}
\label{dates}
\begin{tabular*}{0.95\columnwidth}{l @{\extracolsep\fill} c c c }
\hline
\\[-2ex]
Year & Start date & End date & Number of \\[-1ex]
& (dd.month) & (dd.month) & measurement days \\
\hline
\\[-2ex]
2013 & 08.08 & 22.11 & 102 \\
2014 & 28.04 & 24.10 & 180 \\
2015 & 05.05 & 24.11 & 219 \\
2016 & 06.05 & 03.11 & 187 \\
2017 & 12.04 & 21.11 & 229 \\
2018 & 04.05 & 05.11 & 187 \\
2019 & 15.05 & 22.11 & 192 \\
2020 & 06.05 & 25.10 & 171 \\
\hline
\\
\end{tabular*}

\vspace*{-0.5em}
In total, measurements on 1467 days are considered in the analysis.
\end{table}

For the special analysis the spectra of global radiation in case of
clear sky are extracted for the sun zenith angles of 40\degree,
45\degree, 50\degree, 55\degree, and 60\degree at 530 wavelengths in the
range of 300.1 to 2160.9 nm. The total number of such spectra in every
summer is 291, 400, 526, 649, and 721, respectively, altogether 2587
spectra.  The clear sky spectra were extracted manually, observing that
the global irradiance was stable during several minutes before and after
the moment when the sun zenith angle was equal to one from the
prescribed set (40\degree, 45\degree\ etc).


\section{Results}
\label{results}

\subsection{Global radiation}
\label{global}

\begin{figure}[!ht]
\centering
\resizebox*{0.95\columnwidth}{!}{\includegraphics{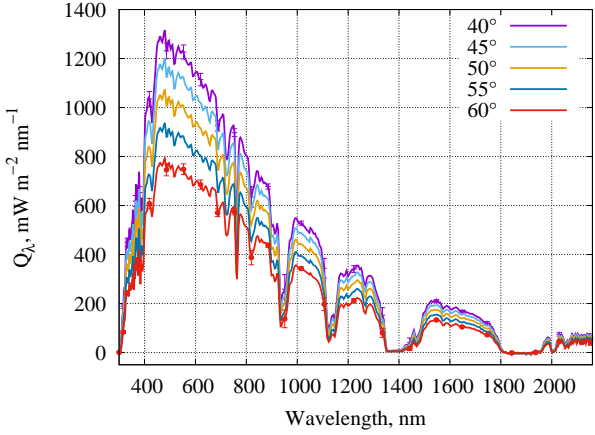}}
\caption{Mean spectra of global radiation in case of clear sky at various
sun zenith angles.}
\label{meanq}
\end{figure}

Mean spectra of global radiation in case of clear sky at various sun
zenith angle are plotted in Fig.  \ref{meanq}. Standard deviations are
marked by errorbars for sun zenith angles of 40\degree\  and 60\degree.

\citet{Kondratyev1965} suggested an approximate solution of the radiative
transfer equation for the calculation of spectral hemispherical solar
radiation during cloudless sky:
\begin{equation}
Q_\lambda(\theta_s) = \frac{S_0(\lambda) \,\mu_s}
{1 + f(\lambda) \,/\, \mu_s} ,
\label{flmbda}
\end{equation}

\noindent where $\theta_s$ is the sun zenith angle, $S_0(\lambda)$ is
the spectral solar constant, $\mu = \cos(\theta_s)$, and $f(\lambda)$ is
a turbidity parameter of the atmosphere. The Solar Irradiance Reference
Spectrum SAO2010 by \citet{Harvard2010} was used as the
extra-terrestrial solar spectrum.

\begin{figure}[!ht]
\centering
\resizebox*{0.9\columnwidth}{!}{\includegraphics{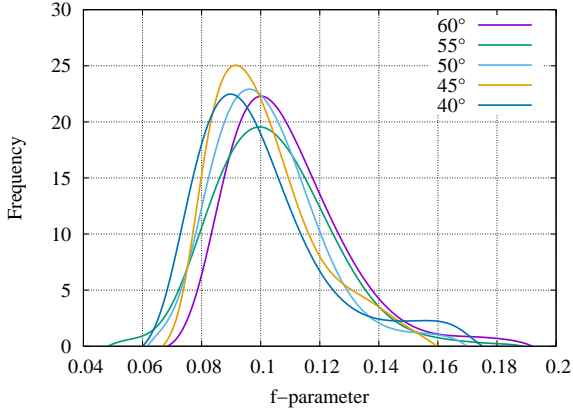}}
\caption{Distribution of $f(549.8)$ for various sun zenith angles.}
\label{fb}
\end{figure}

The approximation Eq.~(\ref{flmbda}) is applicable at wavelengths of low
absorption. The parameter $f(549.8)$ of the 2587 spectra varied between
0.0546 and 0.1864, while most of $f$-values are between 0.08 and 0.13.
Smoothed distribution of the $f$-parameter is plotted in Fig.  \ref{fb}.

\begin{figure}[!ht]
\centering
\resizebox*{0.9\columnwidth}{!}{\includegraphics{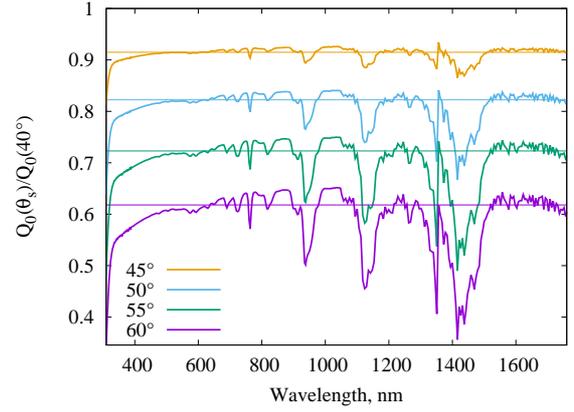}}
\caption{Ratio of mean spectra at various sun zenith angles to the mean
spectrum at $\theta_s = 40\degree$.}
\label{ratio}
\end{figure}

Mean spectra in Fig.~\ref{meanq} are rather similar but in differing
scale. In Fig.~\ref{ratio} the ratio
$Q_\lambda(\theta_s)/Q_\lambda(\theta_s=40\degree)$ is plotted for
different sun zenith angles $\theta_s$. Thin horizontal lines of same
color mark the ratio according to the Eq.~(\ref{flmbda}) for $f(\lambda)
= 0.09$.

There are spectral intervals where this ratio is close to the
theoretical value according to Eq.~(\ref{flmbda}). Deviations of
measured ratios from straight lines in Fig.~\ref{ratio} indicate the
change in spectral composition of global radiation with changing sun
zenith angle. Deviations are large in absorption bands. A systematic
change is the decrease of the share of blue radiation with increasing
sun zenith angle. Fig.~\ref{ratio} provides the quantitative measure for
this change.

\begin{figure}[!ht]
\centering
\resizebox*{0.9\columnwidth}{!}{\includegraphics{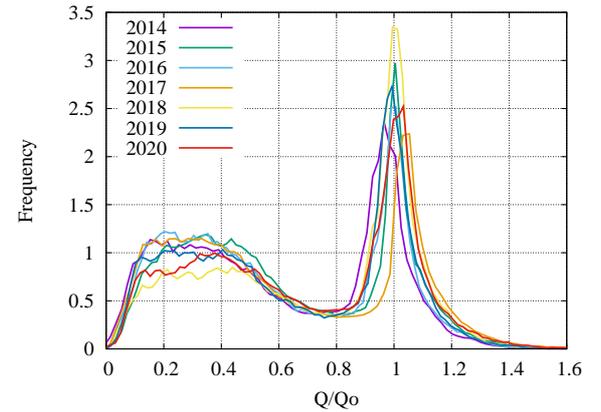}}
\caption{Distribution of $Q/Q_0$ in different years, $Q$ is the measured
total radiation in the spectral range 320--1800~nm, $Q_0$ is calculated
with Eq.~(\ref{flmbda}) using $f = 0.15$.}
\label{qrelat015}
\end{figure}

Principal component analysis of clear sky spectra in the spectral range
300--2160~nm for SZA from 40\degree\ to 60\degree\ revealed that the
first eigenvector describes more than 98.6\% of spectral variability.
Consequently the main change in the solar spectrum is the change of
irradiance level with changing turbidity, and changes in spectral
composition are small. This conclusion and small deviations of measured
curves from straight lines in Fig.~\ref{ratio} in rather wide spectral
intervals allow to apply the approximation Eq.~(\ref{flmbda}) for
integrated solar irradiance in some rather wide spectral range as well.
In Fig.~\ref{qrelat015} the distribution of relative global radiation
$Q/Q_0$ for the whole measurement period (1467 days) in the wavelength
range 320--1800~nm for years from 2014 to 2020 is plotted. The short
measurement period of 2013 is not included in this plot. As the used
spectral range includes absorption bands, the f-parameter value for the
whole spectral range $f(\Delta\lambda) = 0.15$ was used for all years.
In case of wrong $f$-value the peak in Fig.~\ref{qrelat015} moves away
from the value $Q/Q_0 = 1$.

The main shape of the distribution of $Q/Q_0$ is formed by cloudiness.
In general, cloudiness decreases the global solar irradiance compared to
the irradiance in case of no clouds at the same sun zenith angle.  Ratio
values \,$Q/Q_0 > 1$\, are caused by focusing of sun radiation by broken
clouds, mainly by partial cover of Cu-clouds when direct sun radiation
between clouds is not attenuated, and bright clouds near sun increase
the diffuse sky irradiance substantially.

Variations of turbidity modify the peak of distribution at \,$Q/Q_0 =
1$.  Small variations in the position of the peak may be caused both by
different turbidity of the atmosphere in different years and by errors
in the calibration of SkySpec.  The summer of 2018 was very sunny, thus
the peak at \,$Q/Q_0 = 1$\, is higher and the level of the distribution
at $Q/Q_0$ in the range between 0.1 and 0.5 lower than in other years.

\begin{figure}[!ht]
\centering
\resizebox*{0.9\columnwidth}{!}{\includegraphics{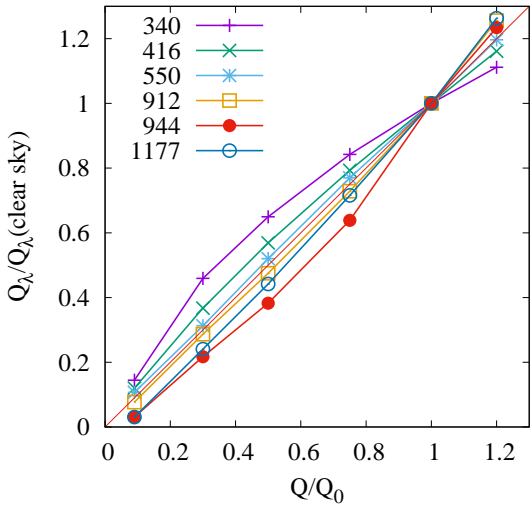}}
\vspace*{-2em}
\caption{The change of share of spectral irradiance in the integrated
irradiance at different wavelengths, sun zenith angle between 
40\degree\ and 50\degree.}
\label{qtoq0_40}
\end{figure}

Change of cloudiness modifies not only the relative global irradiance
$Q/Q_0$ but also the distribution of energy in the spectrum of total
radiation.  With increasing cloudiness the share of spectral irradiance
at some wavelengths increases but at some other wavelength decreases.
Fig.~\ref{qtoq0_40} shows how change of cloudiness (change of relative
global irradiance $Q/Q_0$) modifies relative share of radiation at
various wavelengths in the integral irradiance. The selection of
wavelengths in Fig. \ref{qtoq0_40} includes boh the bands of high
absorption and the wavelengths of no absorption.

\subsection{Diffuse radiation}
\label{diffuse}

Energetic values of spectral diffuse sky radiation vary in wide range.
Here we analyze the relative spectra of diffuse radiation as the ratio
to the spectral global irradiance. The records of global and diffuse
spectra are not simultaneous, therefore the spectral global irradiance
at the moment of recording diffuse irradiance is calculated as the mean
value of the last global spectrum before and the first one after the
diffuse irradiance measurement stage.

\begin{figure}[!ht]
\centering
\resizebox*{0.9\columnwidth}{!}{\includegraphics{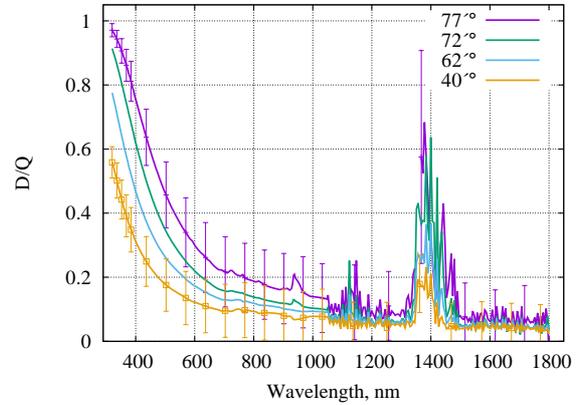}}
\caption{Ratio of diffuse to global spectral irradiance at various sun
zenith angles during clear sky. Error bars mark standard deviation.}
\label{dq100}
\end{figure}

Ratio of diffuse spectral irradiance to global irradiance during clear
sky for various sun zenith angle is plotted in Fig.~\ref{dq100}.  Only
data of 2015 are used in this figure.  The selection of spectra was
based on the ratio $Q/Q_0, ~~ 0.95 < Q/Q_0 < 1.05$, see
Fig.~\ref{qrelat015}. In bands of water absorption at $\lambda =
1130$~nm and $\lambda = 1395$~nm the signal/noise ratio is low and the
$D(\lambda)/Q(\lambda)$ ratio values in Fig.~\ref{dq100} are not
reliable. Error bars mark standard deviation of spectra.

\begin{figure}[!ht]
\centering
\resizebox*{0.9\columnwidth}{!}{\includegraphics{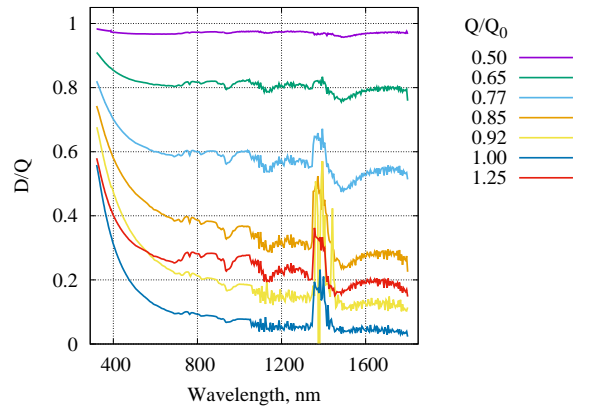}}
\caption{Ratio of diffuse to global spectral irradiance at various
$Q/Q_0$ values, sun zenith angle 40\degree.}
\label{dqxxx_40}
\end{figure}

Figure~\ref{dqxxx_40} shows the $D/Q$ spectra at various $Q/Q_0$ values.
Increase of cloud amount (decrease of $Q/Q_0$ ratio) increases the
general level of diffuse radiation, and also the balance of blue and
NIR/SWIR radiation in diffuse irradiance. In case of $Q/Q_0 > 1$ the
share of red and NIR-SWIR radiation increases substantially.  Ratio
values \,$Q/Q_0 > 1$\, are caused by focusing of sun radiation by
clouds. The exact ratio value depends on the cloud pattern near sun
direction and dependent on weather conditions (wind speed at cloud
level) may change rather fast. That will cause changes in the spectrum
of diffuse sky irradiance as well.

\section{Discussion and conclusions}
\label{summary}

Systematic spectral measurements of downwelling solar radiation, both
global and diffuse, have been collected during 8 years in the
hemi-boreal zone in south east Estonia near the SMEAR-Estonia research
station.  The measurements provided information about the variation of
spectral and total downwelling hemispherical global and diffuse solar
irradiance in the wavelength range from 300 to 2160~nm with spectral
resolution of 3~nm in UV to 16~nm in SWIR spectral regions. Unique data
have been collected and quantitative description of the variability of
the measured spectra is provided. It is well known that clouds modify
substantially irradiance, however there is no single cloudscreening
method which were widely accepted as a standard \citep{Ruizarias2018}.
The difficulty of identifying clear-sky periods a posteriori leads to
likely incorrect comparisons between clear-sky irradiance predictions
and observations under partly cloudy conditions. In this work the ratio
$Q/Q_0$ in the spectral range of 320--1800~nm is used for the
description of the synoptic situation during measurements.  We suggest
to use this measure as the primary description of the weather situation
instead of cloud cover. Cloud cover is estimated in meteorological
stations mainly visually, and thus may have subjective errors. At the
same time the measurement of integrated global radiation with
pyranometers is not very complex nor expensive. The number of
meteostations where the measurements of downwelling solar radiation with
pyranometers are carried out is increasing \citep{Ohmura1998}. Recording
of solar irradiance is much more simple than the instrumental
measurement of cloud cover at a common meteorological station.

Collected data can be used for the validation of atmospheric radiative
transfer models. Collected data allowed to estimate variable atmospheric
parameters for the atmospheric correction of satellite images
\citep{Kuusk2018b}.

Spectra of global and diffuse irradiance for the period and spectral
range of interest are available at Tartu Observatory, Estonia. Please
contact authors for arranging details.

\section*{Acknowledgements}
This work was supported by Estonian Environmental Observatory, Project
3.2.0304.11-0395, by the EU Regional Development Fund, and by Estonian
Research Council grants PUT~232 and PUT~1355.


\end{document}